\documentclass{article}

\usepackage{PRIMEarxiv}

\usepackage[dvipsnames]{xcolor}

\usepackage[utf8]{inputenc} 
\usepackage[T1]{fontenc}    
\usepackage{hyperref}       
\usepackage{url}            
\usepackage{booktabs}       
\usepackage{amsfonts}       
\usepackage{nicefrac}       
\usepackage{microtype}      
\usepackage{xcolor}         
\usepackage{amsmath}
\usepackage{amsthm}
\usepackage{amssymb}
\usepackage{amsfonts}
\usepackage{graphicx}
\usepackage{subcaption}   
 \usepackage{bbm}
 \usepackage{tabularx}
 \usepackage{wrapfig}
 \usepackage{nameref}
 \usepackage[round]{natbib}
\title{6DCNN with roto-translational convolution filters for volumetric data processing
}
\author{
  Dmitrii Zhemchuzhnikov, Ilia Igashov, and Sergei Grudinin \\
  Univ. Grenoble Alpes, CNRS, Grenoble INP, LJK \\
  38000 Grenoble, France \\
  \texttt{\{dmitrii.zhemchuzhnikov,ilia.igashov,sergei.grudinin\}@univ-grenoble-alpes.fr} \\
}

\begin{document}
\maketitle

\begin{abstract}
In this work, we introduce 6D Convolutional Neural Network (6DCNN) designed to tackle the problem of detecting relative positions and orientations of local patterns when processing three-dimensional volumetric data. 
6DCNN also includes 
SE(3)-equivariant
 message-passing and nonlinear activation operations constructed in the Fourier space. 
Working in the Fourier space allows significantly reducing the computational complexity of our operations. 
We demonstrate the properties of the 6D convolution and its efficiency in the recognition of spatial patterns. 
We also assess the 6DCNN model on several datasets from the recent CASP protein structure prediction challenges. 
Here, 6DCNN
improves over the baseline architecture and also outperforms the state of the art.
\end{abstract}


\section{Introduction}
Methods of deep learning have recently made a great leap forward in the spatial data processing. 
This domain contains various tasks from different areas of industry and natural sciences, including three-dimensional (3D) data analysis.
 For a long time, convolutional neural networks (CNNs) remained the main tool in this domain. 
 CNNs helped to solve many real-world challenges, especially in computer vision. 
 However, these architectures have rather strict application restrictions.
 Unfortunately, real-world raw data rarely have standard orientation and size, which limits the efficiency of translational convolutions. 
 This circumstance has created an increased interest in the topic of SE(3)-equivariant operations 
 in recent years. 

Moreover, volumetric data may contain arbitrarily-positioned and -oriented sub-patterns.
Their detection makes 3D pattern recognition particularly challenging.
Indeed, a volumetric sub-pattern in 3D has six degrees of freedom (DOFs), three to define a rotation, and three for a translation.
Thus, 
a classical convolution technique would require scanning through all these six DOFs and scale as $O(N^3M^6)$ if a brute-force computation is used,
where $N$ is the linear size of the volumetric data, and $M$ is the linear size of the sub-pattern.

This work proposes a set of novel 
operations with the corresponding architecture
based on six-dimensional (6D) {\em roto-translational} convolutional filters.
For the first time, thanks to the polynomial expansions in the Fourier space, 
we demonstrate the feasibility of the 6D roto-translational-based convolutional network
with the leading complexity of $O(N^2 M^4)$ operations.
We tested our method on simulated data and also on protein structure prediction datasets, 
where the overall accuracy of our predictions is on par with the state-of-the-art methods.
Proteins play a crucial role in most biological processes. 
Despite their seeming complexity, structures of proteins attract more and more attention from the data science community \citep{senior2020improved,Jumper_2021,laine2021protein}.
In particular, the task of protein structure prediction and analysis raises  the challenge of constructing rotational and translational equivariant architectures.

\section{State of the art / Related work}

	\textbf{Equivariant operations.}
	The first attempt of learning rotation-equivariant representations was made in Harmonic Networks \citep{worrall2017harmonic} in application to 2D images. 
	Further, this idea was transferred to the 3D space with the corresponding architecture known as 3D Steerable CNNs \citep{weiler20183d}. 
	In Spherical CNNs \citep{cohen2018spherical}, the authors introduced a correlation on the rotation group and proposed a concept of rotation-equivariant CNNs on a sphere. 
	Spherical harmonics kernels that provide rotational invariance
	have also been applied to point-cloud data \citep{poulenard2019effective}.
	A further effort on leveraging compact group representations resulted in the range of methods based on Clebsh-Gordan coefficients \citep{kondor2018n, kondor2018clebsch, anderson2019cormorant}. 
	This approach was finally generalized in Tensor field networks \citep{thomas2018tensor}, where rotation-equivariant operations were applied to vector and tensor fields. 
	Later, SE(3)-Transformers \citep{fuchs2020se} were proposed to efficiently capture the distant spatial relationships.
	More recently, \cite{hutchinson2021lietransformer} continued to develop the theory of equivariant convolution operations for homogeneous spaces and proposed Lie group equivariant transformers, following works on the general theory of group equivariant operations in SO(2) \citep{romero2020attentive,romero2021group} and SO(3) \citep{cohen2020general}. 
	We should mention that the most common data representation in this domain is a 3D point cloud, however, several approaches operate on regular 3D grids \citep{weiler20183d, pages2019protein}.
    We should also add that some of the above-mentioned methods \citep{cohen2018spherical, weiler20183d, kondor2018n, anderson2019cormorant} employ Fourier transform in order to learn rotation-equivariant representations.

\textbf{Geometric learning on molecules}. 
	As graphs and point clouds are natural structures for representing molecules, it is reasonable that geometric learning methods have been actively evolving especially in application to biology, chemistry, and physics. 
	More classical graph-learning methods for molecules include MPNNs \citep{gilmer2017neural}, SchNet \citep{schutt2017schnet}, and MEGNet \citep{chen2019graph}.
	Ideas for efficient capturing of spatial relations in molecules resulted in rotation-invariant message-passing methods DimeNet \citep{klicpera2020directional} and DimeNet++ \citep{klicpera2020fast}. Extending message-passing mechanism with rotationally equivariant representations, polarizable atom interaction neural networks \citep{schutt2021equivariant} managed to efficiently predict tensorial properties of molecules. \cite{satorras2021n} proposed E(n) equivariant GNNs for predicting molecular properties and later used them for developing generative models equivariant to Euclidean symmetries \citep{satorras2021n2}.
	
	\textbf{Proteins} are much bigger and more complex systems than small molecules but are composed of repeating blocks.
	Therefore, more efficient and slightly different methods are required to operate on them. 
	A very good example is the two recent and very powerful methods AlphaFold2 \citep{Jumper_2021} and RoseTTAFold \citep{baek2021accurate}.
	Most recent geometric learning methods designed for proteins include deep convolutional networks processing either volumetric data in local coordinate frames \citep{pages2019protein, hiranuma2021improved},
	graph neural networks \citep{ingraham2021generative, sanyal2020proteingcn, baldassarre2021graphqa, Igashov:2021vf, igashov2021spherical}, deep learning methods on surfaces and point clouds \citep{gainza2020deciphering, sverrisson2020fast}, and geometric vector perceptrons \citep{jing2021learning,jing2021equivariant}. 
	In addition, several attempts were made to scale tensor-field-like  SE(3)-equivariant methods to proteins \citep{derevyanko2019protein,eismann2020protein,townshend2020geometric,baek2021accurate}.

\section{Model/Method}

\subsection{Workflow}
\begin{figure}[htbp]
  \centering
  \includegraphics[width=1\textwidth]{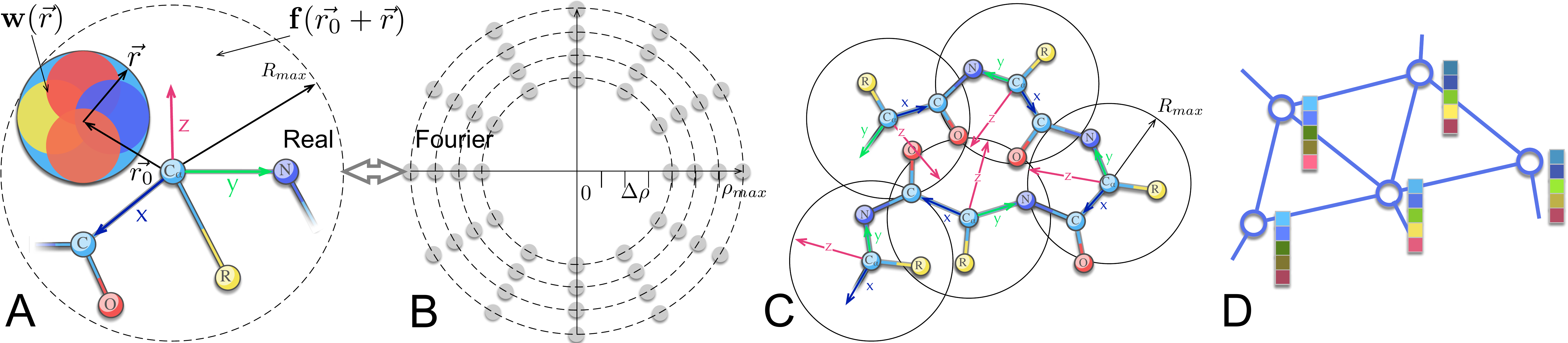}
   \caption{{\bf A}. Six-dimensional (6D) convolution between a filter $\mathbf{w}(\vec{r})$ and a function $\mathbf{f}(\vec{r_0}+\vec{r})$.
   The function $\mathbf{f}(\vec{r_0}+\vec{r})$ describes the local environment of a protein residue and is defined within a certain radius $R_{max}$ from the corresponding $C_{\alpha}$ atom. 
   The local coordinate system $xyz$ is built on the backbone atoms  $C_{\alpha}$,  $C$, $N$ of each protein residue.
   $R$ denotes the location of a residue's side-chain.
   {\bf B}. The spherical Fourier space with the reciprocal spacing $\Delta \rho$ and the maximum resolution of $\rho_{max}$. 
   Grey dots schematically illustrate points where the Fourier image is stored.
   {\bf C}. An illustration of a protein chain representation. 
   Each protein residue has its own coordinate system $xyz$ and the corresponding local volumetric description $\mathbf{F}^{k}_{l} (\rho)$ within a certain sphere of $R_{max}$ radius.
   Spheres of different residues may overlap. Two residues are considered as neighbors in the graph representation if their  $C_{\alpha}$ atoms are located within 
a certain threshold  $R_{n}$. 
 {\bf D}. The graph representation of the protein structure. The node features are learned by the network and are represented with colored rectangles.
 The edge features are assigned based on the types of the corresponding residues and the topological distance of the protein graph. }
 \label{fig:geom}
  \end{figure}

Here, we give a brief description of all steps of our method that are described in more detail below. 
Firstly, for each residue in the input protein molecule, we construct a function $\mathbf{f}(\vec{r})$ that describes its local environment.
More technically, 
this function is a set of 3D Gaussian-shaped features centered on the location of atoms within a certain distance  $R_{max}$ from the $C_{\alpha}$ atom of the corresponding residue (see Fig. \ref{fig:geom}A).

Then, for each function $\mathbf{f}(\vec{r})$, we compute its spherical Fourier expansion coefficients $\mathbf{F}^{k}_{l} (\rho)$.
The angular resolution of the expansion is determined by the maximum order of spherical harmonics $L$.
The radial resolution of the expansion corresponds to the maximum reciprocal distance $\rho_{max}$ and is inversely proportional to the resolution $\sigma$ of the real-space Gaussian features as
$\rho_{max} = \pi / \sigma$ (see Fig. \ref{fig:geom}B).
Similarly, the radial spacing between the reciprocal points is inversely proportional to the linear size of the data, 
$\Delta \rho =  \pi / (2R_{max})$.
Without loss of generality, we can set the number of reciprocal radial points to be equal $L$, such that $\rho_{max} / \Delta \rho = L+1 = 2R_{max} / \sigma$.

Spherical Fourier coefficients  $\mathbf{F}^{k}_{l} (\rho)$
constitute the input for our network, along with the information about the transition from the coordinate system of one residue to another.
We start the network with the embedding block that reduces the dimensionality of the feature space.
Then, we apply a series of 6D convolution blocks that consist of 6D convolution, normalization, and activation layers, followed by a message-passing layer.
After a series of operations on continuous data, we switch to the discrete representation and continue the network with graph convolutional layers (see Fig. \ref{fig:geom}C-D).
In the graph representation, each node corresponds to a protein residue, and a graph edge links nodes if the distance between the corresponding  $C_{\alpha}$ atoms is smaller than a certain threshold $R_n$.
We should also mention that the backbone structure of a protein residue can be 
used to unambiguously define its local coordinate system \citep{pages2019protein,Jumper_2021} using the Gram–Schmidt orthogonalization process
starting from  $C_{\alpha}-N$ and $C_{\alpha}-C$ vectors.

\subsection{Representation of volumetric data}
Let us consider a function $\mathbf{f}(\vec{r}): \mathcal{R}^3 \rightarrow \mathcal{R}^{d_f}$ that describes a distribution of $d_f$-dimensional features in the 3D space. 
Very often, initial data is given as a point cloud, as it is typically the case for protein structures.
Let us consider a set of points located within a maximum radius $R_{max}$ at positions  $\vec{r}_1, ..., \vec{r}_n, ..., \vec{r}_N$ with the corresponding feature vectors $\mathbf{t}_1, ..., \mathbf{t}_n, ..., \mathbf{t}_N$. 
To convert this representation to a continuous one, we assume that each point feature has a Gaussian shape with a standard deviation $\sigma$.
Then, the continuous function characterizing this area will have the form,
\begin{equation}
\mathbf{f}(\vec{r}) = \sum_{n = 1}^{N}  \mathbf{f}_n (\vec{r})= \sum_{n = 1}^{N}  \mathbf{t}_n \exp(-\frac{(\vec{r} - \vec{r}_n)}{2 \sigma^2}),
\end{equation}
where $\sigma$ is the {\em spatial resolution} of the features.
It is very convenient to use the Fourier description of this function in spherical coordinates.
The spherical harmonics expansion of the Fourier transform of function $ \mathbf{f} (\vec{r})$ will be
\begin{equation}
\mathbf{F}^{k}_{l} (\rho) = 4 \pi (-i)^l    \sum_{n = 1}^{N}  j_l(\rho r_n) \overline{Y_{l}^{k}(\Omega_{r_n})} (\sqrt{2 \pi} \sigma)^3  \exp(-\frac{\sigma^2 \rho^2}{2 }) \mathbf{t}_n,
\end{equation} 
where $\rho$ is the {\em reciprocal distance}.
Please see Appendices \ref{app:math}-\ref{app:fc_gf} for more details.

\subsection{6D convolution operation}
The initial idea that prompted us to consider non-traditional types of convolution is the intention to perceive spatial patterns in whatever orientations they have. 
In a classical 3D convolution, when a filter is applied to the image, the result is sensitive to the orientation of the filter. 
Keeping this in mind, we came with the idea to extend the convolution with an integration of all possible filter rotations. 
Let $\mathbf{f}(\vec{r}):  \mathcal{R}^3 \rightarrow \mathcal{R}^{d_i}$ and $\mathbf{w}(\vec{r}): \mathcal{R}^3 \rightarrow \mathcal{R}^{d_i} \times  \mathcal{R}^{d_o}$  be the initial signal and a spatial filter, correspondingly. 
We propose to extend the classical convolution as follows,
\begin{equation}
	\int_{\vec{r}}  d \vec{r} \ \mathbf{f}(\vec{r_0} + \vec{r}) \mathbf{w}(\vec{r})  \rightarrow \int_{\Lambda} d \Lambda\  \int_{\vec{r}} d \vec{r}  \ \mathbf{f}(\vec{r_0} + \Lambda^{-1} \vec{r}) \mathbf{w}(\Lambda \vec{r}) ,
	\label{eq:corr}
\end{equation}
where $\Lambda \in \text{SO}(3)$ is a 3D rotation. 
Please see Fig. \ref{fig:geom}A for an illustration.
Let the functions $\mathbf{f}(\vec{r})$ and $\mathbf{w}(\vec{r})$
be {\em finite-resolution} and
have 
spherical Fourier expansion coefficients  $F_l^k(\rho)$ and $W_l^k(\rho)$, correspondingly, 
which are nonzero for $l \leq$ than some maximum expansion coefficient $L$.
Then, the result of the 6D convolution has the following coefficients,
\begin{equation}
   [\mathbf{F}_{\text{out}}]_{l}^{k} (\rho) =  \sum_{l_1 = 0}^{L} \sum_{k_1 = -l_1}^{l_1} \frac{8 \pi^2}{2 l_1 +1}  \mathbf{W}_{l_1}^{-k_1} (\rho)  \sum_{l_2 = | l - l_1|}^{l + l_1}  c^{l_2} (l, k, l_1 , -k_1) \mathbf{F}_{l_2}^{k+k_1} (\rho),
    \label{eq:conv6d_coeffs}
\end{equation}
where $c^{l}$ are the products of three spherical harmonics, see Eq. \ref{eq:slater}. 
The proof can be found in Appendix \ref{app:proof}.        
For  a single reciprocal distance $\rho$,
the complexity of this operation is $O(L^5)$, where $L$ is the maximum order of the spherical harmonics expansion.

We should specifically draw the reader's attention to the fact of opposite rotations $\Lambda$ and $\Lambda^{-1}$  in the convolution equation \ref{eq:corr}. 
Indeed, if we leave only one rotation on $\mathbf{f}(\vec{r})$ or $\mathbf{w}(\vec{r})$, after the integration over rotations in SO(3), we will obtain  the resultant convolution of the input function that will depend only on the radial components of the filter.
Alternatively, if we apply the same rotation on both $\mathbf{f}(\vec{r})$ and $\mathbf{w}(\vec{r})$, it will be integrated out, and we will obtain a common 3D translational convolution as a result.

\subsection{Nonparametric  message passing of continuous data}

Let us assume that our spatial data 
can be represented with overlapping spatial fragments, each having its own local coordinate frame, as it is shown in Fig. \ref{fig:geom}C for the continuous representation of a protein molecule. 
Then, we can recompute the fragment representation in the neighboring coordinate frames using spatial transformation operators. 
For this purpose, we designed a message passing operation in a form that recomputes the spherical Fourier coefficients in a new reference frame.
We should specifically note that this operation is SE(3)-equivariant by construction, 
because the spatial relationship between molecular fragments remain the same when rotating and shifting the global coordinate system.
We decompose such a spatial transformation in a sequence of a rotation followed by a $z$-axis translation followed by the second rotation. 
Indeed, the spherical Fourier basis provides low computational complexity for the $z$-translation and rotation operations. 
Let function $\mathbf{f}_z(\vec{r})$ be the results of translating function $\mathbf{f}(\vec{r})$ along the $z$-axis by an amount $\Delta$.
Then, the expansion coefficients of these two functions will have the following relation,

\begin{equation}
[\mathbf{F}_z]_{l}^{k} (\rho)  =  \sum_{l' = k}^{L} T_{l, l'}^{k} (\rho , \Delta) \mathbf{F}_{l'}^{k} (\rho) + O\left(\frac{1}{(L-l)!}(\frac{ \rho  \Delta}{2})^{L-l}\right),
\label{eq:shift_coefs}
\end{equation}
where $T_{l, l'}^{k}$ is a translation tensor specified in Appendix \ref{app:transl_oper}.
The update of all the expansion coefficients costs $O(L^4)$ operations.

Similarly, let function $\mathbf{f}_{\Lambda}(\vec{r})$ be the rotation of function $\mathbf{f}(\vec{r})$ by an amount $\Lambda \in \text{SO(3)}$, $\mathbf{f}_{\Lambda}(\vec{r}) = \mathbf{f}(\Lambda\vec{r})$.
The corresponding expansion coefficients are then related as
\begin{equation}
[\mathbf{F}_{\Lambda}]_{l}^{k} (\rho) = \sum_{k' = -l}^{l} D_{k', k}^{l} (\Lambda) \mathbf{F}_{l}^{k'} (\rho),
\label{eq:rot_coefs}
\end{equation}
where  $D_{k', k}^{l}$ is a rotation Wigner matrix specified in Appendix \ref{app:rot_oper}.
The update of all the expansion coefficients will again cost $O(L^4)$ operations.

\subsection{Normalization}
Since we are working with continuous data in the Fourier space, we have to introduce our own activation and normalization functions. 
When developing the normalization, we proceeded from very basic premises. 
More precisely, we normalize  the  signal $\mathbf{f}(\vec{r})$ by setting its mean to zero and its variance to unity.
This can be achieved if the following operations are performed on the spherical Fourier expansion coefficients of the initial function,

\begin{equation}
[\mathbf{F}_n]^k_l(\rho) = 
\begin{cases}
0, & \text{ if } l=k=\rho =0, \\
[\mathbf{F}]^k_l(\rho)/\mathbf{S_2} , & \text{otherwise},
\end{cases}
\label{eq:norm_coefs}
\end{equation}

where $\mathbf{S_1} = \int_{\mathcal{R}^3} \mathbf{f}(\vec{r})  d \vec{r}$,  $\mathbf{S_2} = \int_{\mathcal{R}^3} (\mathbf{f}(\vec{r}) - \mathbf{S_1})^2 d \vec{r}$.
We should also notice that we apply the element-wise division in Eq. \ref{eq:norm_coefs}.
The proof can be found in Appendix \ref{app:normalization}.

\subsection{Activation}
The concept of our activation operation coincides with the idea of the classical activation in neural networks, 
i.e. to nonlinearly transform the initial signal depending on how it differs from the bias. 
We aimed to extend this idea  to the continuous case. 
Let the initial signal be $\mathbf{f}(\vec{r})$, and the bias signal with trainable Fourier coefficients be $\mathbf{b}(\vec{r})$.
Then, we propose the following output of the normalization-activation block,
\begin{equation}
\mathbf{f}_a(\vec{r}) = \left(\frac{1}{4}\int_{\mathbb{R}^3}(N(\mathbf{f}(\vec{r}) + \mathbf{b}(\vec{r})) - N(\mathbf{b}(\vec{r})))^2 d^3 \vec{r}\right) N(\mathbf{f}(\vec{r}) + \mathbf{b}(\vec{r})),
\end{equation}
where $N()$ is the normalization operation defined in Eq. \ref{eq:norm_coefs} such that the value $\frac{1}{4}\int_{\mathbb{R}^3}(N(\mathbf{f}(\vec{r}) + \mathbf{b}(\vec{r})) - N(\mathbf{b}(\vec{r})))^2 d^3 \vec{r}$  lies in the interval $[0, 1]$.
If the signal $\mathbf{f}(\vec{r})$ is 'coherent' to $\mathbf{b}(\vec{r})$ , $\mathbf{f}(\vec{r}) = K\mathbf{b}(\vec{r}), K > 0$, then it does not pass the block. 
The amplification factor reaches its maximum value when the two signals are anti-coherent. 
Parseval's theorem allows us to translate these formulas into operations on their expansion coefficients,
\begin{equation}
[\mathbf{F}_a]^k_l(\rho) = \frac{1}{4} \left(\sum_{l'=0}^{L} \sum_{k' = -l'}^{l'} \int_{0}^{\infty} ( N(\mathbf{F}^{k'}_{l'}(\rho) + \mathbf{B}^{k'}_{l'}(\rho)) - N(\mathbf{B}^{k'}_{l'}(\rho)))^2  \rho^{2} d \rho \right)  N(\mathbf{F}^k_l(\rho) + \mathbf{B}^k_l(\rho)),
\label{eq:activation_coefs}
\end{equation}
where $[\mathbf{F}_a]^k_l(\rho_p) $  and $\mathbf{B}^k_l(\rho_p)$ are the expansion coefficients of functions $\mathbf{f}(\vec{r})$ and $\mathbf{b}(\vec{r})$, correspondingly.

\subsection{Switching from the continuous to the vector representation}
For most of the real-world tasks that may require the proposed architecture, a transition from a continuous functional  representation $\mathbf{f}(\vec{r}):  \mathcal{R}^3 \rightarrow \mathcal{R}^{d_f}$ to a discrete vector representation $\mathbf{h} \in \mathcal{R}^{d_f}$ is necessary. 
There can be several ways to achieve it.
In the simplest case, 
when the input function $\mathbf{f}(\vec{r})$ can be unambiguously associated with some reference coordinate system,
as in the case of protein's peptide chain, we may use the following operation,
\begin{equation}
	\mathbf{h} = \int_{\mathcal{R}^3} \mathbf{f}(\vec{r}) \mathbf{w}(\vec{r}) d \vec{r},
\end{equation}
where $\mathbf{w}(\vec{r}):  \mathcal{R}^3 \rightarrow \mathcal{R}^{d_f}$ is a filter function element-wise multiplied by the input function.
If the functions $\mathbf{f}(\vec{r})$ and $\mathbf{w}(\vec{r})$ have corresponding expansion coefficients $\mathbf{F}^k_l(\rho_p) $ and $\mathbf{W}^k_l(\rho_p) $, then this operation will have the following form (please see more details in Appendix \ref{app:overlap}),
\begin{equation}
	\mathbf{h} =\sum_{l=0}^L
	\sum_{k=-l}^{l}
	 \int_{0}^{\infty} 
	 \mathbf{F}^k_l(\rho) \overline{\mathbf{W}}^k_l(\rho)
	 \rho^2 d \rho.
	\label{eq:cont_to_discrete_coefs}
\end{equation}

\subsection{Graph convolutions layers}
In our model, the continuous representation is followed by the classical graph convolutional layers (see Fig.  \ref{fig:geom}C-D). 
Indeed, protein structures, on which we assess our model, allow us to use the graph representation. 
In such a graph, each node corresponds to an amino acid residue and characterizes the 3D structure of its neighborhood,
and each edge between two nodes
indicates their spatial proximity, i.e. the distance between the corresponding C-alpha atoms within a certain threshold $R_{n}$ (see Fig. \ref{fig:geom}D).

Let us consider a graph $\mathcal{G}$   that is described by the feature matrix $\mathbf{H} \in \mathcal{R}^{N \times d_v}$, where $N$ is the number of graph nodes and $d_v$ is the dimensionality of the node feature space, and the adjacency matrix $ \mathbf{A} \in \mathcal{R}^{N \times N \times d_e}$, where $d_e$ is the dimensionality of the edge feature space. 
We decided to use one-hot edge features that would encode the types of amino acids of the associated nodes.
To reduce the dimensionality of the edge feature space to $d_e'$, we use the following trainable  embedding  $\mathbf{R}_e \in \mathcal{R}^{d_e \times d_e'}$, 
and a reduced adjacency matrix $\mathbf{A}_r = \mathbf{A} \mathbf{R}_e$.
Finally, the graph convolution step is defined as
\begin{equation}
\mathbf{H}_{k+1} = \sigma_a(\mathbf{A}_r  \mathbf{H}_{k} \mathbf{W}_k + \mathbf{H}_{k} \mathbf{W}_k^s + \mathbf{b}_k) ,
\label{eq:gcl}
\end{equation}
where $\mathbf{W}_k \in \mathcal{R}^{d_{k} \times d_{k+1} \times d_{e}'}$ and $\mathbf{W}^s_k \in \mathcal{R}^{d_{k} \times d_{k+1}}$ are trainable matrices.

\section{Experiments}
\subsection {6D filters}

\begin{figure}[htbp]
  \centering
  \includegraphics[width=1\textwidth]{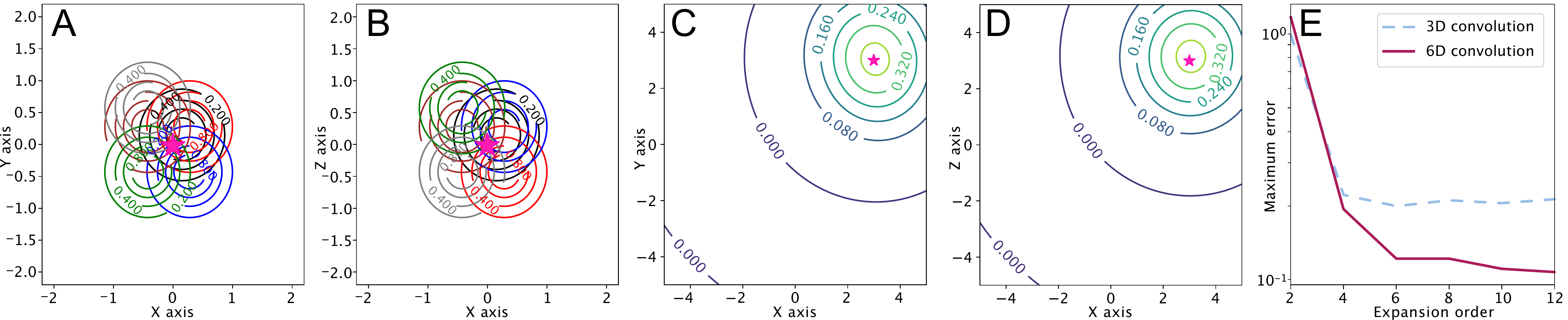}
   \caption{
   {\bf A-B.} Six Gaussian volumetric features of $\sigma=0.4$ \AA~shown in the $xy$-plane (A) and $xz$-plane (B).
   The center of mass of the whole pattern is shown with the pink star.
   {\bf C-D.}
   The density map of the resultant 6D convolution between an original pattern and its translated and rotated copy to the pink star position,
   shown in the $xy$-plane (C) and $xz$-plane (D).
{\bf E.}
   The maximum error in determining the center of the translated and rotated pattern as a function of the expansion order $L$ divided by the Gaussian feature $\sigma$ for
  3D and 6D convolutions.
  }
 \label{fig1}
  \end{figure}

Our first step was to study the properties of the 6D roto-translational convolution operation. 
To do so, we generated a small 3D pattern,  $\mathbf{f}(\vec{r})$, composed of six 3D Gaussians with $\sigma=0.4$ \AA~shown in Fig. \ref{fig1}A-B.
We then created a function $\mathbf{h}(\vec{r})$, rotated and translated $\mathbf{f}(\vec{r})$ to a new location.
Figure \ref{fig1}C-D shows the result of the 6D convolution between  $\mathbf{h}(\vec{r}) $ and $\mathbf{f}(\vec{r})$  given by  Eq. \ref{eq:conv6d_coeffs}.
As we can expect, the maximum of this convolution corresponds to the position of the center of mass of function   $\mathbf{h}(\vec{r})$,
and the value of the convolution reduces as we go further from this point.
We used the following parameters for this experiment,
$\sigma=0.4$ \AA,
 $L = 4$, 
$\rho_{\max}  = 0.6 \pi$ \AA$^{-1}$, 
$\Delta \rho  = 0.2 \pi$ \AA$^{-1}$.

We then compared the result of the classical 3D convolution (see Appendix \ref{3DFourier}) with the proposed 6D convolution. 
Using the same initial pattern, randomly rotated multiple times, we recorded the maximum positional error in determining the center of mass of the translated pattern 
with respect to the maximum expansion order $L$.
As we can see in Fig. \ref{fig1}E, the 6D convolution detects the position of the shifted pattern more accurately compared to its classical 3D counterpart, and the accuracy increases with 
the maximum expansion order $L$.
  
\subsection{Message passing and translation operator}

\begin{figure}[htbp]
  \centering
    \includegraphics[width=1\textwidth]{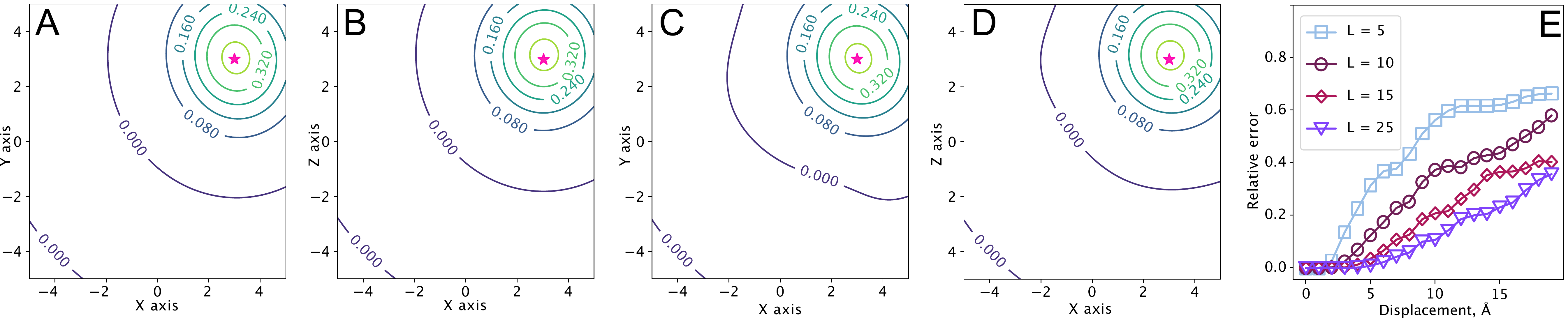}

    \caption{
    {\bf A-B.} Results of the 6D convolution in the original coordinate system, shown in the $xy$-plane (A) and $xz$-plane (B).
    {\bf C-D.} Results of the 6D convolution in the translated coordinate system, by $3\sqrt{3}$ \AA, shown in the $xy$-plane (C) and $xz$-plane (D).
    {\bf E.} Relative error of the recovering  volumetric patterns as a function of the translation amplitude and the expansion order.}
    \label{fig2}
  
\end{figure}

Our next experiment was the assessment of the message-passing step.
Our main goal was to study the conditions of validity for the translation operator \ref{eq:shift_coefs}, which is also implicitly used in the 6D convolution part.
Specifically, we were interested in whether the result of the 6D convolution will be preserved after changing the coordinate systems.
For this experiment, we used the same volumetric pattern  $\mathbf{f}(\vec{r})$ described above and shown in Fig. \ref{fig1}A-B.
We then shifted this pattern to a new location and recorded the value of the 6D convolution, as described above, shown in  Fig. \ref{fig2}A-B.
For the comparison, we computed the same 6D convolution from a different coordinated system, shifted by  $3\sqrt{3}$ \AA~from the original one.
We used parameters from the previous experiment.
The result is shown in  Fig. \ref{fig2}C-D.
Both convolution functions have their maximums near the location of the center of mass of the shifted pattern, however, their volumetric shapes are slightly different.

For a more rigorous experiment, we examined the relative error of the translation operator \ref{eq:shift_coefs} as a function of the displacement of the coordinate system and the expansion order.
Here, we fixed parameters to
$\sigma=0.4$ \AA,
$\rho_{\max}  = 0.6 \pi$ \AA$^{-1}$, 
$\Delta \rho  = 0.2 \pi$ \AA$^{-1}$,
and varied the value of $L$.
From the results shown in Fig.  \ref{fig2}E we can see that 
for the displacements within about $\sigma L/2$,
the error is negligibly small, which is the consequence of the Nyquist–Shannon theorem.

\subsection {Datasets}
In order to train and assess our models, we used data from the Critical Assessment of Structure Prediction (CASP)
benchmarks \citep{Kryshtafovych:2019tr,Moult:1995wu}. 
Each model from the benchmarks has a quality score that indicates how close it is to the experimentally determined solution. 
There are several quality metrics designed for protein models, e.g., GDT-TS  \citep{Zemla:1999tw}, CAD-score \citep{Olechnovic:2013wq}, or lDDT \citep{Mariani:2013tc}.  
We  trained our networks to predict the local per-residue lDDT score, which shows the precision of the local 3D structure and topology around each amino acid.
We also considered the average value of predicted local scores as the global score for the model.

We used models from CASP8-11 stage 2 submissions as the training  and validation sets. 
We randomly split the targets into two sets.
The training set contained $301$ target structures with $104,985$ models. 
The validation set contained $32$ target structures with $10,872$ models. 
Finally, we tested the networks on CASP12 stage2 dataset  that has  $39$ native structures  and $5,462$ models, and also on CASP13 stage2 dataset with $18$ native structures and $2,583$ models. 
All the sets included only the models for which the native structure is publicly available.

\subsection {Evaluation metrics}
We assessed our networks on both global and local lDDt scores. 
In order to evaluate the efficiency of our networks, we engaged the following metrics: z-scores, determination coefficient  $R^2$, and Pearson and Spearman correlations. 
Z-scores are equal to the ground-truth score of the top-predicted model divided by the standard deviation of the models for a particular target.
We computed $R^2$ and correlations in two modes, global and per-target. The per-target approach assumed calculating metrics for each target separately and averaging results over all the targets. 
We used  the Fisher transformation to extract average values. 
For the global approach, we calculated metrics for all model structures of regardless which target they belong to. 
For the local structure quality predictions, as CASP submissions do not explicitly store local lDDt scores,
we assessed our models using Spearman rank correlation of the submitted distance deviations with the ground-truth data.

\subsection{Loss function}
Our loss function contains two terms, the first 
is the MSE between the predictions and the ground-truth local scores, and the second is the MSE of the z-scores between these two arrays,
\begin{equation}
L(s_i, p_i) = \alpha \frac{1}{N}\sum_{i =1}^{N} (s_i - p_i)^2   +(1-\alpha) \sigma_s^2 \frac{1}{N}\sum_{i =1}^{N} (\frac{s_i - \mu_s}{\sigma_s} - \frac{p_i - \mu_p}{\sigma_p})^2,
\end{equation}
where $N$ is the number of residues in the protein model; 
$\mu_s = \frac{1}{N}\sum_{i}^{N} s_i$,   $\sigma_s = \sqrt{\frac{1}{N}\sum_{i}^{N} (s_i -\mu_s)^2} $; $\mu_p$ and $\sigma_p$ have the same formulas but for the array $p_i$. 
To make two terms in the loss function to have approximately the same amplitude,
we introduced an  additional variance of  the ground-truth local scores factor $\sigma_s^2$.
The weight $\alpha$ is a hyperparameter. 
We need both terms in the loss function to simultaneously increase the correlation of the local predictions and to also predict the absolute values of the scores.

\subsection{Technical details}

Amino-acid residues can contain atoms of $167$ distinguishable types.
We have also included one additional type for the solvent molecules.
Overall, the dimensionality of the volumetric function characterizing the protein model $N_a = 168$.
We choose some maximum value of the radius  $R_{{max}}$,
which limits the set of atoms that fall in the amino-residue neighborhood.
We also choose some parameter $R_n$,
which is the maximum distance between amino residues that considered as neighbors in the graph.

We use $20$ amino acids types.
The edge between two amino acids can be determined by a vector of the size $20 \times 20 + d_t$, 
where the first $20 \times 20$ elements are a one-hot representation of the amino acids pair, with the distinguishable order of residues in a pair. 
The last $d_t$ elements in this vector is a one-hot representation of the topological distance between the residues in the protein graph.
 The values of $R_{{max}}, R_{n}, d_{t}$, and $L$ are the hyper-parameters of the network that were optimized with a grid search.
 
 During the training process, 
 we preprocessed all data using C++ code 
 before feeding them to the network. 
 For each residue of each protein model, we precomputed its Fourier representation.
 In the case of the architectures with the message-passing step, 
 we have also generated tensors for the transformation of Fourier coefficients between coordinate system frames along the edges of the graph.  

\subsection {Baseline architecture}
\label{section:baseline}
For the comparison, we introduced a baseline architecture that would help us to assess the novel layers.
It begins with trainable embedding in the feature space. 
We then applied the transition from the continuous to the discrete representation using operation \ref{eq:cont_to_discrete_coefs} that is followed by three graph convolutional layers described in Eq. \ref{eq:gcl}. 
For the activation, we used the $\tanh$ function in the last layer and the LReLU function with the 'leak'  parameter of $0.05$ in all other layers. 
We also introduced two trainable parameters $\mu_t$ and $\sigma_t$ for the mean and the standard deviation of the local rates  in the training sample. 
The relationship between the output of the last layer $l_N$ and the output of the network $o$ is $o = \sigma_t l_N + \mu_t$. 
Overall, the baseline architecture had 21,026 trainable parameters.

{\textbf{Training.}} This network was trained on CASP 8-11 datasets in $1,280$ iterations. 
At each iteration, the network was fed by $16$ input protein models. 
One training iteration took $\approx 5$ minutes on Intel \textcopyright Xeon(R) CPU E5-2630 v4 @ 2.20GHz.

{\textbf{Hyperparameters.}} The network has the following hyperparameters, $\sigma = 2$ \AA, $R_n = 12$ \AA, $d_{t} = 10$, $L = 4$, and $R_{{max}} = 8 $ \AA.

\subsection {6DCNN networks}
\label{section:6dcnn}
The main difference between the baseline architecture and the 6DCNN networks is the presence of 6D convolution layers in the latter. 
Our first architecture (6DCNN-1) contains only one 6D convolution layer. 
The second architecture (6DCNN-2) has two 6DCNN layers. 
The 6DCNN layer is composed of the following operations: 6D convolution, followed by normalization and activation. 
Consecutive  6D convolution layers are linked with the message passing step.
Table \ref{table1} lists the architectures of the networks.
Overall, the 6DCNN-1 and 6DCNN-2 architectures had  187,326 and 283,426 trainable parameters, correspondingly.
Figure \ref{fig:Hists_learncurv_filters}A shows real-space projections of two 6D convolution filters learned by 6DCNN-1.

{\textbf{Training.}} The two networks were trained on CASP 8-11 datasets in $1,280$ iterations. 
At each iteration, the networks were fed by $16$ protein models. 
One training iteration took $\approx 6$ minutes for 6DCNN-1 and $\approx 15$ minutes  for 6DCNN-2 on Intel \textcopyright Xeon(R) CPU E5-2630 v4 @ 2.20GHz.
Figure \ref{fig:Hists_learncurv_filters}D demonstrates the learning curves on the validation dataset of three architectures, baseline, 6DCNN-1, and 6DCNN-2. 

{\textbf{Hyperparameters.}} 
The networks have the following hyperparameters, $\sigma = 2$ \AA, $R_n = 12$ \AA, $d_{t} = 10$, $L = 4$, and $R_{{max}} = 8 $ \AA.

\subsection {CASP results}
In order to assess the 6DCNN architectures, 
we compared their performance on the CASP12 (Table \ref{table2}) and CASP13  (Table \ref{table3}) datasets with the baseline model and also with the state-of-the-art single-model quality assessment methods SBROD, SVMQA, VoroCNN, Ornate, ProQ3, and VoroMQA  \citep{Cheng:2019uh}. 
SBROD is a linear regression model that uses geometric features of the protein backbone \citep{Karasikov:2019wy}.
SVMQA is a support-vector-machine-based method that also uses structural features  \citep{Manavalan:2017uh}. 
VoroMQA engages statistical features of 3D Voronoi tessellation of the protein structure \citep{Olechnovic:2017tv}. 
VoroCNN is a graph neural network built on the  3D Voronoi tessellation  of protein structures \citep{Igashov:2021vf}.
Ornate is a convolutional neural network that uses 3D volumetric representation of protein residues in their local reference frames \citep{pages2019protein}. 
ProQ3 is  a fully connected neural network operating on the precomputed descriptors \citep{Uziela:2016tf}.
We computed the ground-truth lDDT values ourselves. 
Therefore, we were forced to limit the datasets to only those models that had publicly available target structures. 
As a result, the CASP12 dataset  turned out to be significantly bigger than CASP13, with more demonstrative and representative results.

\begin{figure}[htbp]
  \centering
   
    \includegraphics[width=1\textwidth]{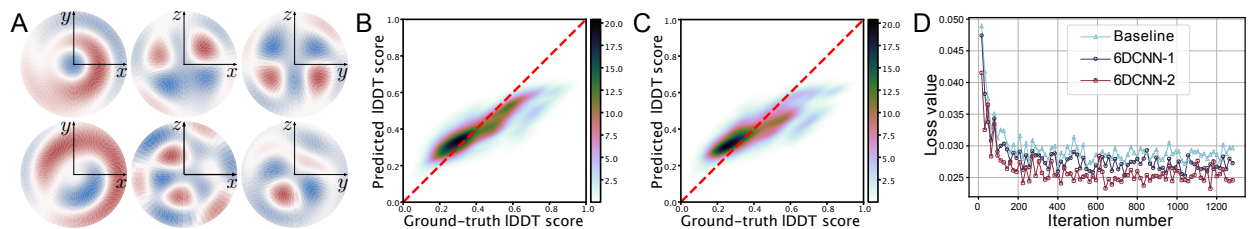}
    \caption{
    {\bf A.} Examples of  two filters learned by  6DCNN-1, two rows represent the two filters. Each column demonstrates different filter projections. 
	The red color represents the positive filter values, and the blue color corresponds to the negative values.
   {\bf B-C.} Density maps comparing the ground-truth global lDDT scores with the predictions of the 6DCNN networks on the CASP12 dataset, 6DCNN-2 (left, B) and 6DCNN-1 (right, C).
    {\bf D.} Loss function on the validation dataset during the training of the 6DCNN networks and the baseline model.   }
    \label{fig:Hists_learncurv_filters}
\end{figure}

On the CASP12 test set, we achieved a noticeable improvement in comparison with the state-of-the-art methods.
Even though the difference between the 6DCNN networks and the baseline model performance is not big,
one of the 6DCNN architectures outperforms the baseline in  every metric except for the z-score.
We can also notice that the 6DCNN-2 method gives significantly higher global correlations  and $R^2$ metrics on the CASP12 dataset than 6DCNN-1 and all other methods.
However, 6DCNN-1 demonstrates better per-target correlations on CASP12 data  than 6DCNN-2.
Both of the networks have higher per-target correlations than most of the state-of-the-art methods.
Unfortunately, we did not manage to achieve satisfying performance on the z-score metric.
However, z-scores are rather noisy compared to correlations, and not directly linked to the optimized loss function. 
The fact that 6DCNN-2 has better global correlation scores 
confirms the importance of the additional 6D correlation block.
Figure \ref{fig:Hists_learncurv_filters} (B-C) shows correlations between the ground-truth global scores from the CASP12 dataset and the corresponding predictions by the two 6DCNN models. 
The 6DCNN-2 map has a higher density near the diagonal, indicating a better absolute predictions of global scores and a better $R^2$ metric.

On the CASP13 dataset, we did not greatly outperform the state-of-the-art methods (see Table \ref{table3}).
However, we reached a performance that is on par with the state of the art. 
Moreover, we should notice that 6DCNN-2 did not outperform  6DCNN-1. 
This can be explained by the fact that we trained our models on CASP[8-11] datasets, which are rather different from CASP13,
and also that the CASP13 dataset is less representative than CASP12.

Table \ref{table:local_scores} from Appendix \ref{app:local_scores} lists Spearman rank correlations of local quality predictions with the corresponding ground-truth values of our networks and the state-of-the-art methods on model structures of $11$ targets from CASP13.  
For the comparison, we chose only those models that had both publicly available target structures and local score predictions by all other methods. 
As we did not have these predictions for the CASP12 dataset, we limited local score evaluation by CASP13 data only.
Here, we did not achieve the  best results that could be explained by the small size of the dataset.

\begin{table}[ht]
\centering
\begin{tabular}{ | c | c  ccc c ccc |} 
\hline
 & & \multicolumn{3}{c}{Global} & &\multicolumn{3}{c}{Per-target} \\
 \cline{3-5} \cline{7-9}
 Method & z-score & $R^2$ & Pearson, $r$ &Spearman, $\rho$& &$R^2$ & Pearson, $r$ & Spearman, $\rho$ \\
\hline
SBROD & 1,29& -33,66  & 0,55 & 0,54 &&-325,18 &0,76&  0,67\\
SVMQA	& $\mathbf{1,48}$&	0,32	&0,82&	0,80	&&-2,44&	0,76	& $\mathbf{0,73}$\\
VoroCNN&	1,39	&0,61&	0,80&	0,80&&	-2,79&	0,73&	0,69 \\
Ornate&	1,42&	0,36	&0,78	&0,77&&	-5,14	&0,73 &	0,69 \\
ProQ3&	1,18 &	0,26 &	0,74 &	0,77 && 	-4,85 &	0,73 &	0,68 \\
VoroMQA &	1,18 &	-0,25 & 	0,59 & 	0,62 &&	-6,23 &	0,74 &	0,70 \\
\hline
\hline
Baseline & 	1,34 & 	0,55 &	0,82 &	0,82 &&	-2,35 &	0,78 &	0,71 \\
6DCNN-1	& 1,26 & 	0,57 & 0,81 & 	0,81&& 	-2,29 &$\mathbf{0,80}$ &	$0,73$ \\
6DCNN-2	 & 1,19& 	$\mathbf{0,63}$ &	$\mathbf{0,85}$ &	$\mathbf{0,84}$ &&	$\mathbf{-1,95}$ &	0,79 &	0,71 \\
\hline

\end{tabular}
\caption{Comparison of the 6DCNN networks with the baseline architecture and the state-of-the-art methods on the unrefined CASP12 stage2 dataset.
The best value for each metric is highlighted in bold.
}
\label{table2}
\end{table}

\begin{table}[ht]
\centering
\begin{tabular}{ | c | c  ccc c ccc |} 
\hline
 & & \multicolumn{3}{c}{Global} & &\multicolumn{3}{c}{Per-target} \\
 \cline{3-5} \cline{7-9}
 Method & z-score & $R^2$ & Pearson, $r$ &Spearman, $\rho$& &$R^2$ & Pearson, $r$ & Spearman, $\rho$ \\
\hline

SBROD &	1,23 &	0,07 & 0,72 & 0,69 &&	-1,44 & 	0,81 &	0,74 \\
VoroCNN &1,15 &	$\mathbf{0,67} $&	$\mathbf{0,84}$ & $\mathbf{0,82}$ && 	0,03 &	0,79 &	0,77 \\
VoroMQA & 1,32 & 0,20 &	0,77 &	0,79 &&	-1,10 &	0,79 & 	0,75 \\
ProQ3D & $\mathbf{1,39}$&	-0,03 &	0,75 &	0,75 && -2,07 & 0,76 &	0,72\\
Ornate & $0.93$&0,26 &	0.62 &	0.67 && -2,10 & 0,78 &	0,77\\
\hline
\hline
Baseline &  1,35& 0,44&  0,78& 0,77&& -0,30& 0,82&  0,77   \\
6DCNN-1	& 1,03 & 	0,59 &	0,82 &	0,80 &&	-0,02 &	$\mathbf{0,83}$ &		$\mathbf{0,79}$ \\
6DCNN-2	 & 1,30	& 0,56 &	0,79 &	0,78 &&	-0,12&	0,82&	0,77 \\
\hline

\end{tabular}
\caption{Comparison of the 6DCNN networks with the baseline architecture and the state-of-the-art methods on the unrefined CASP13 stage2 dataset.
The best value for each metric is highlighted in bold.
}
\label{table3}

\end{table}

\section{Conclusion}

This work presents a theoretical 
foundation for 6D roto-translational spatial patterns detection and 
the construction of neural network architectures for learning on spatial continuous data in 3D.
We built several networks that consisted of 6DCNN blocks followed by GCNN layers specifically designed for 3D models of protein structures.
We then tested them on the CASP datasets from the community-wide protein structure prediction challenge.
Our results demonstrate that 6DCNN blocks are able to accurately learn local spatial patterns and improve the quality prediction of protein models. 
The current network architecture can be extended in multiple directions, for example, including the attention mechanism.



\section*{Appendix}
\appendix

\section{Mathematical prerequisites}
\label{app:math}
\subsection{Spherical harmonics  and spherical Bessel transform}
Spherical harmonics are complex functions defined on the surface of a sphere that constitute a complete set of orthogonal functions and thus an orthonormal basis.  
They are generally defined as
\begin{equation}
Y^k_l(\Omega) = \sqrt{{\frac{(2l+1)(l-k)!}{4 \pi (l+k)!}}} P_l^k(\cos(\psi_{\Omega})) \exp{i k \theta_{\Omega}},
\end{equation}  
where $0 \leq \psi_{\Omega} \leq \pi$ is the colatitude of the point $\Omega$, and $0 \leq \theta_{\Omega} \leq 2\pi$  is the longitude. 
As mentioned above, these functions are orthonormal,
\begin{equation}
\int_{4 \pi} Y^k_l(\Omega)  \overline{Y^{k'}_{l'}}(\Omega) d \Omega = \delta_{l l'} \delta_{k k'},
\end{equation}
where the second function $ \overline{Y^{k'}_{l'}}(\Omega) $ is complex-conjugated.
An expansion of a function $f(\vec{r})$ in spherical harmonics has the following form,
\begin{equation}
f_l^k (r) = \int_{4 \pi} f(\vec{r})  \overline{Y^k_l(\Omega_r)} d \Omega_r.
\end{equation}
Spherical Bessel transform (SBT, sometimes referred to as spherical Hankel transform)  of order $l$ computes Fourier coefficients of spherically symmetric functions in 3D,
\begin{equation}
{F}_l(\rho) = \text{SBT}_l(f(r)) =  \int f(r) j_l(\rho r) r^2 dr,
\end{equation}
where $j_l(r)$ are spherical Bessel functions of order $l$.   
The inverse transform has the following form,
\begin{equation}
f(r) = \text{SBT}_l^{-1}({F}_l(\rho))= \frac{2}{\pi} \int {F}_l(\rho) j_l(\rho r) \rho^2 d \rho.
\end{equation}

Spherical Bessel functions of the same order are orthogonal with respect to the argument,
\begin{equation}
\int_{0}^{\infty} j_l(\rho_1 r) j_l(\rho_2 r) r^2 d r = \frac{\pi}{2 \rho_1 \rho_2} \delta(\rho_1 - \rho_2).
\end{equation}

\subsection{Plane wave expansion}
The plane wave expansion  is the decomposition of  a plane wave to a linear combination of spherical waves.  
According to the spherical harmonic addition theorem, 
a plane wave can be expressed through spherical harmonics and spherical Bessel functions,
\begin{equation}
\exp{i \vec{\rho} \vec{r}} = 4 \pi \sum_{l = 0}^{\infty} \sum_{k=-l}^{l}  i^l j_l(\rho r ) Y_{l}^{k} (\Omega_{\rho})  \overline{Y_{l}^{k} (\Omega_{r})}.
\label{eq:pwe}
\end{equation}

\section{3D Fourier transforms in spherical coordinates \label{3DFourier}}
The 3D Fourier transform of a function $f(\vec{r})$ is defined as
\begin{equation}
F(\vec{\rho}) = \int_{\mathbb{R}^3}  f(\vec{r}) \exp^{i \vec{r} \vec{\rho}} d \vec{r}.
\end{equation}
The spherical harmonics expansion of this transform has the following form,
\begin{equation}
F_l^k(\rho) = \int_{4 \pi} F(\vec{\rho})  Y_{l}^{k} (\Omega_{\rho}) d \Omega_{\rho}.
\end{equation}
The Fourier spherical harmonics expansion coefficients relate to the real-space spherical harmonics coefficients though SBT,
\begin{equation}
F_l^k(\rho) =  4 \pi (-i)^{l} \text{SBT}_l(f_l^k(r)) .
\end{equation}
This equation can be also rewritten as follows,
\begin{equation}
F_l^k(\rho) =  4 \pi (-i)^{l} \int_{\mathbb{R}^3} f(\vec{r})  j_l(r \rho) \overline{Y^k_l(\Omega_r)} d \vec{r}.
\end{equation}

\section{Rotation of spherical harmonics and Wigner $D$-matrices}
\label{app:rot_oper}
The Wigner $D$-matrices $\mathbf{D}_l$ are the irreducible representations of $\text{SO}(3)$ that can be applied  to spherical harmonics functions 
to express the rotated functions with tensor operations on the original ones,
\begin{equation}
Y_l^k(\Lambda \Omega) = \sum_{k'=-l}^{l} D_{ k k'}^{l} (\Lambda) Y_l^{k'}(\Omega),
\end{equation}
where $\Lambda \in \text{SO}(3)$.

\section{Translation operator}
\label{app:transl_oper}
Let us translate a 3D function  $f(\vec r)$ along the $z$-axis by an amount $\Delta$. 
The expansion coefficients of a translated function will be
\begin{equation}
[F_z]_l^{k}(\rho) = \int F(\vec{\rho})e^{-i\Delta\vec{\rho}.\vec{e_z}}\overline{Y_{l}^{k}(\Omega)}d\Omega.
\end{equation}
Using the plane-wave expansion 
and triple spherical harmonics integrals defined though Slater coefficients  $c^{l_2}(l, k, l_1, k_1)$,
\begin{equation}
c^{l_2}(l, k, l_1, k_1) = \int_{4 \pi} \overline{Y_l^k(\Omega)} Y_{l_1}^{k_1}(\Omega) Y_{l_2}^{k -k_1}(\Omega) d \Omega,
\label{eq:slater}
\end{equation}
we obtain
\begin{equation}
\begin{split}
[F_z]_l^k(\rho) &= \sum_{p=0}^{\infty} \sum_{l' = \max(|l-p|,|k|)}^{l+p} 
i^p
j_{p}(\rho \Delta)
F_{l'}^k(\rho)
4\pi \sqrt{\frac{2p+1}{4\pi}}
\int_{4 \pi}Y_l^k(\Omega)Y_p^0(\Omega)
\overline{Y_{l'}^{k}(\Omega)} d\Omega \\
&=
\sum_{p=0}^{\infty} \sum_{l' = \max(|l-p|,|k|)}^{l+p} 
i^p
j_{p}(\rho \Delta)
F_{l'}^k(\rho)
4\pi \sqrt{\frac{2p+1}{4\pi}}
c^{p}(l',k,l,k). 
\end{split}
\end{equation}
Changing the summation order and introducing the maximum expansion order $L$, we arrive at
\begin{equation}
[{F}_z]_{l}^{k} (\rho)  =  \sum_{l' = |k|}^{L} T_{l, l'}^{k} (\rho , \Delta) F_{l'}^{k} (\rho) + O\left(\frac{1}{(L-l)!}(\frac{ \rho  \Delta}{2})^{L-l}\right),
\label{eq:shift_coefs}
\end{equation}
where
\begin{equation}
 T_{l, l'}^{k} (\rho , \Delta) =
\sum_p 
 i^p
j_{p}(\rho \Delta)
4\pi \sqrt{\frac{2p+1}{4\pi}}
c^{p}(l',k,l,k). 
\label{eq:finite_res_transl_operator}
\end{equation}

The error estimation in Eq. \ref{eq:shift_coefs} is based on the equivalence relation $j_p(x) \sim \frac{1}{\sqrt{xp}} (\frac{ex}{2p})^p$ as $p \rightarrow \infty$ that allows us to find the upper bound of the error,
\begin{equation}
\sum_{p=L+1}^{\infty} \sum_{l' = \max(|l-p|,|k|)}^{l+p} 
i^p
j_{p}(\rho \Delta)
F_{l'}^k(\rho)
4\pi \sqrt{\frac{2p+1}{4\pi}}
c^{p}(l',k,l,k) \leq C \frac{(\frac{e\rho \Delta}{2})^2}{(L-l-1)!},
\end{equation}
where $C$ is a constant independent of $L$.

\section{Fourier coefficients for a 3D Gaussian function}
\label{app:fc_gf}
Let us consider a function of the Gaussian type in 3D,
\begin{equation}
    f(\vec{r}) = \exp( - \frac{(\vec{r} - \vec{r}_0)^2}{2 \sigma^2}).
\end{equation}
The Fourier transform of $f(\vec{r})$ will be
\begin{equation}
    \begin{split}
        F(\vec{\rho}) = \int_{\mathcal{R}^3} \exp( - \frac{(\vec{r} - \vec{r}_0)^2}{2 \sigma^2}) \exp( -i\vec{r}\vec{\rho}) d^3 \vec{r} \\ = \exp(-\frac{i 2 \sigma^2 \vec{r}_0 \vec{\rho} +\sigma^4 \rho^2}{2 \sigma^2}) \int \exp(- \frac{r^2 +2\vec{r}(i\sigma^2\vec{\rho} - \vec{r}_0) + r_0^2 -2 i \sigma^2 \vec{r}_0 \vec{\rho} - \sigma^4 \rho^2}{2 \sigma^2}) d^3 \vec{r} \\ = \exp(-i   \vec{r}_0 \vec{\rho}) \exp(-\frac{\sigma^2 \rho^2}{2 }) (\sqrt{2 \pi} \sigma)^3.
    \end{split}
     \label{eq:fc_gf2}
\end{equation}
Using the  plane wave expansion \eqref{eq:pwe},  we obtain
\begin{equation}
    F(\vec{\rho}) = 4 \pi \exp(-\frac{\sigma^2 \rho^2}{2 }) (\sqrt{2 \pi} \sigma)^3 \sum_{l = 0}^{\infty} \sum_{k = -l}^{l}  (-i)^l j_l(\rho r_0) \overline{Y_{l}^{k}(\Omega_{r_0})} Y_{l}^{k}(\Omega_{\rho}) = \sum_{l = 0}^{\infty} \sum_{k = -l}^{l}  F_{l}^{k}(\rho) Y_{l}^{k}(\Omega_{\rho}).
     \label{eq:fc_gf3}
\end{equation}
Consequently,
\begin{equation}
    F_l^{k}(\rho) = 4 \pi (-i)^l  \exp(-\frac{\sigma^2 \rho^2}{2 }) (\sqrt{2 \pi} \sigma)^3  j_l(\rho r_0) \overline{Y_{l}^{k}(\Omega_{r_0})} .
     \label{eq:fc_gf4}
\end{equation}

\section{Proof  of  Eq. \eqref{eq:conv6d_coeffs} }
\label{app:proof}
\begin{proof}
We start in the proof from the translation operator in 3D \citep{baddour20103dfourier},
\begin{equation}
\begin{split}
f(\vec{r} - \vec{r}_0) = \sum_{l = 0}^{\infty} \sum_{k = -l}^{l} 8 (i)^l Y^k_l(\Omega_{r_0}) \sum_{l_1 = 0}^{\infty} \sum_{k_1 = -l_1}^{l_1} (-i)^{l_1}\overline{Y_{l_1}^{k_1} (\Omega_{r})} \\ \sum_{l_2 = \|l-l_1\|}^{l+l_1} (-i)^{l_2} c^{l_2} (l,k,l_1, k_1) \int_0^{\infty} f_{l_2}^{k - k_1} (u) S_{l_2}^{l,l_1} (u, r_0, r) u^2 d u,
\end{split}
\end{equation}
where $c^{l_2}(l, k, l_1, k_1)$ are Slater coefficients defined in Eq. \ref{eq:slater}.
They are nonzero only for $\|l -l_1\| \leq l_2 \leq \|l +l_1\| $. 
$S_{l_2}^{l,l_1} (u, r, r_0) $ is a triple Bessel product which is defined as
\begin{equation}
S_{l_2}^{l,l_1} (u, r, r_0) = \int_{0}^{\infty} j_{l_2} (\rho u)  j_{l_1} (\rho r_0)  j_{l} (\rho r) \rho^2 d \rho
\end{equation} 
Let us consider the function  $f(\vec{r_0} - \Lambda^{-1} \vec{r})$, where $ \Lambda \in \text{SO}(3)$,
\begin{equation}
\begin{split}
    f(\vec{r_0} - \Lambda^{-1} \vec{r}) = \sum_{l_1 = 0}^{\infty} \sum_{k_2 = -l_1}^{l_1} 8 (i)^{l_1} Y_{l_1}^{k_2} (\Omega_{r_0}) \sum_{l_2 = 0}^{\infty} \sum_{k_3 = -l_2}^{l_2} \sum_{k_4 = -l_2}^{l_2} (-i)^{l_2} \overline{Y_{l2}^{k_3}(\Omega_r)} D_{k_4 k_3}^{l_2}(\Lambda^{-1}) \\ \sum_{l_3 = |l_1 - l_2|}^{l_1 + l_2} (-i)^{l_3} c^{l_3} (l_1, k_2, l_2, k_4) \int_{0}^{\infty} f_{l_3}^{k_2 - k_4} (u) S_{l_3}^{l_1, l_2} (u, r_0, r) u^2 d u  .
\end{split}
\end{equation}
Using the symmetry properties of Wigner $D$-matrices,
\begin{equation}
    D_{k_4 k_3}^{l_2}(\Lambda^{-1}) = (-1)^{(-k_4 + k_3)}D_{-k_3 -k_4}^{l_2}(\Lambda),
\end{equation}
we can retrieve the following equation for the convolution operation in 6D,
\begin{equation}
\begin{split}
    h(\vec{r_0}) = \int_{\vec{r}} \int_{\Lambda} f(\vec{r_0} - \Lambda^{-1} \vec{r}) g(\Lambda \vec{r}) d \vec{r} d \Lambda \\ = \int_{\vec{r}} \int_{\Lambda} \sum_{l = 0}^{\infty} \sum_{k = -l}^{l} \sum_{k_1 = -l}^{l} g_l^{k_1} (r) \overline{D_{k_1 k}^l(\Lambda)} Y_{l}^{k}(\Omega_r)  \sum_{l_1 = 0}^{\infty} \sum_{k_2 = -l_1}^{l_1} 8 (i)^{l_1} Y_{l_1}^{k_2} (\Omega_{r_0})\\ \sum_{l_2 = 0}^{\infty} \sum_{k_3 = -l_2}^{l_2} \sum_{k_4 = -l_2}^{l_2} (-i)^{l_2} \overline{Y_{l_2}^{k_3}(\Omega_r)}(-1)^{(-k_3+k_4)} D_{-k_3 -k_4}^{l_2}(\Lambda) \\ \sum_{l_3 = |l_1 - l_2|}^{l_1 + l_2} (-i)^{l_3} c^{l_3} (l_1, k_2, l_2, k_4) \int_{0}^{\infty} f_{l_3}^{k_2 - k_4} (u) S_{l_3}^{l_1, l_2} (u, r_0, r) u^2 d u  d \vec{r} d \Lambda
    .
\end{split}
\end{equation}
Using the orthogonality of Wigner $D$-matrices,
\begin{equation}
    \int_{\text{SO}(3)} \overline{D_{k_1 k}^l( \Lambda)} D_{k_3 k_2}^{l_1}( \Lambda) d  \Lambda = \frac{8 \pi^2}{2 l+1} \delta_{l l_1} \delta_{k k_2} \delta_{k_1 k_3},
\end{equation}
and the orthogonality of spherical harmonics,
we obtain
\begin{center}
\begin{equation}
\begin{gathered}
    h(\vec{r_0}) = \int_{0}^{\infty} \int_{0}^{\infty} \sum_{l = 0}^{\infty} \sum_{k = -l}^{l} \sum_{k_1 = -l}^{l}  \\ \sum_{l_1 = 0}^{\infty} \sum_{k_2 = - l_1}^{l_1} \sum_{l_2 = 0}^{\infty} \sum_{k_3 = - l_2}^{l_2} \sum_{k_4 = -l_2}^{l_2} \sum_{l_3 = | l_1 - l_2|}^{l_1 + l_2} 8 (i)^{l_1} (-i)^{l_2} (-i)^{l_3} \frac{8 \pi^2}{2 l+1} \delta_{l l_2} \delta_{k_1 -k_3} \delta_{k -k_4} \delta_{k k_3} \delta_{l l_2} \\ Y_{l_1}^{k_2} (\Omega_{r_0}) g_l^{k_1}(r) c^{l_3} (l_1, k_2, l_2, k_4) (-1)^{(k_3-k_4)} f_{l_3}^{k_2 - k_4}(u) S_{l_3}^{l_1, l_2}(u, r_0, r)  u^2 d u r^2 d r \\ =  \int_{0}^{\infty} \int_{0}^{\infty} \sum_{l = 0}^{\infty} \sum_{k = -l}^{l} \sum_{l_1 = 0}^{\infty} \sum_{k_2 = - l_1}^{l_1} \sum_{l_3 = | l_1 - l_2|}^{l_1 + l_2} 8 (-i)^{l} (i)^{l_1} (-i)^{l_3} \frac{8 \pi^2}{2 l +1} g_l^{-k} (r) \\  c^{l_3} (l_1, k_2, l , -k) Y_{l_1}^{k_2}(\Omega_{r_0}) f_{l_3}^{k_2 + k}(u) S_{l_3}^{l_1, l}(u, r_0, r) u^2 d u r^2 d r
    .
\end{gathered}
\end{equation}
\end{center}
Let us  change indices: $l_1 \rightarrow l$, $l \rightarrow l_1$, $k_2 \rightarrow k$, $k \rightarrow k_1$ $l_3 \rightarrow l_2$,
\begin{equation}
\begin{split}
         h(\vec{r_0}) = \sum_{l = 0}^{\infty} \sum_{k = -l}^{l} Y_{l}^{k}(\Omega_{r_0}) 8(i)^l \int_{0}^{\infty} \sum_{l_1 = 0}^{\infty} \sum_{k_1 = -l_1}^{l_1} \frac{8 \pi^2}{2 l_1 +1} (-i)^{l1} \left( \int_{0}^{\infty} g_{l_1}^{-k_1} (r) j_l(\rho r) r^2 d r\right) \\ \sum_{l_2 = | l - l_1|}^{l + l_1} (-i)^{l_2} c^{l_2} (l, k, l_1 , -k) \left( \int_{0}^{\infty} f_{l_2}^{k+k_1} (u) j_{l_2}(\rho u) u^2 d u\right) j_l(\rho r_0) \rho^2 d \rho = \\ \sum_{l = 0}^{\infty} \sum_{k = -l}^{l} Y_{l}^{k}(\Omega_{r_0}) \frac{2}{\pi^2} (i)^l \int_{0}^{\infty} \left( \sum_{l_1 = 0}^{\infty} \sum_{k_1 = -l_1}^{l_1} \frac{8 \pi^2}{2 l_1 +1}  G_{l_1}^{-k_1} (\rho)  \sum_{l_2 = | l - l_1|}^{l + l_1}  c^{l_2} (l, k, l_1 , -k_1) F_{l_2}^{k+k_1} (\rho) \right) j_l(\rho r_0) \rho^2 d \rho.
\end{split}
\end{equation}
Using the spherical harmonics expansion,
\begin{equation}
     h(\vec{r_0}) = \sum_{l = 0}^{\infty} \sum_{k = -l}^{l} Y_{l}^{k}(\Omega_{r_0}) \frac{1}{2 \pi^2} (i)^l \int_{0}^{\infty} H_{l}^{k} (\rho)   j_l(\rho r_0) \rho^2 d \rho,
\end{equation}
we obtain the final result,
\begin{equation}
    H_{l}^{k} (\rho) =  \sum_{l_1 = 0}^{\infty} \sum_{k_1 = -l_1}^{l_1} \frac{8 \pi^2}{2 l_1 +1}  G_{l_1}^{-k_1} (\rho)  \sum_{l_2 = | l - l_1|}^{l + l_1}  c^{l_2} (l, k, l_1 , -k_1) F_{l_2}^{k+k_1} (\rho).
 \end{equation}
 \end{proof}

\section{Normalization}
\label{app:normalization}
Let us consider a function $f(\vec{r}): \mathcal{R}^3 \rightarrow \mathcal{R}$ and its integral over $\mathcal{R}^3$, $ \int_{\mathcal{R}^3} f(\vec{r}) d^3 \vec{r}$.
Using the fact that  $\forall \Omega ,Y_0^{0} (\Omega)  = \frac{1}{\sqrt{4 \pi}} $ and $j_{0} (0) = 1$  and  the properties of orthogonality for  spherical harmonics and spherical Bessel functions we rewrite this expression as
\begin{equation}
\int_{\mathcal{R}^3} f(\vec{r}) d^3 \vec{r}  = \int_{0}^{\infty} \int_{4\pi} \sum_{l = 0}^{\infty} \sum_{k = -l} \frac{(i)^l}{4 \pi} \frac{2}{\pi} \int_{0}^{\infty} F^k_l(\rho) j_l(r \rho) \rho^2 d \rho \sqrt{4 \pi} Y^k_l(\Omega) Y_0^0(\Omega) d \Omega j_0(0 r) r^2 d r =  \frac{1}{\sqrt{4 \pi}} F_0^0(0).
\end{equation}
If we set $F_0^0(0)$ to zero, we obtain the Fourier expansion of  a function $f_m(\vec{r})$ that can be formulated as follows,
\begin{equation}
f_m(\vec{r}) = f(\vec{r}) -\int_{\mathcal{R}^3} f(\vec{r}) d^3 \vec{r} .
\end{equation}
Below we also explain how to normalize the integral of a square of a function to unity using Parseval's theorem.

 \section{Overlap integrals in 3D}
\label{app:overlap}

 \subsection{Parseval's theorem}
Let us consider two finite-resolution complex-valued functions $g(\vec{r})$ and $f(\vec{r})$, meaning that 
their spherical Fourier expansion coefficients
$F_l^k(\rho)$ and  $G_l^k(\rho)$ are nonzero only for $l \leq$ than some maximum expansion order $L$.
Parseval's theorem for these coefficients states the following  \citep{baddour20103dfourier},

 \begin{equation}
 \int_{\mathcal{R}^3} g(\vec{r})\overline{f(\vec{r})} d \vec{r} 
 = \frac{1}{(2 \pi)^3} \sum_{l = 0}^{\infty} \sum_{k = -l}^{l} \int_{0}^{\infty} G_l^k(\rho) \overline{F_l^k(\rho) } \rho^2 d \rho 
 = \frac{1}{(2 \pi)^3} \sum_{l = 0}^{L} \sum_{k = -l}^{l} \int_{0}^{\infty} G_l^k(\rho) \overline{F_l^k(\rho) } \rho^2 d \rho 
,
 \end{equation}
 In the case  $ g(\vec{r}) =  f(\vec{r})$,  it reduces to
  \begin{equation}
 \int_{\mathcal{R}^3} \|f(\vec{r})\|^2 d \vec{r} 
 = \frac{1}{(2 \pi)^3} \sum_{l = 0}^{\infty} \sum_{k = -l}^{l}  \int_{0}^{\infty}  \|F_l^k(\rho_p) \|^2 \rho_p^2 d \rho 
 =  \frac{1}{(2 \pi)^3} \sum_{l = 0}^{L} \sum_{k = -l}^{l}  \int_{0}^{\infty}  \|F_l^k(\rho_p) \|^2 \rho_p^2 d \rho 
.
 \end{equation}

\subsection{Switching from continuous to vector representation}

Let $\mathbf{f}(\vec{r}): \mathcal{R}^3 \rightarrow \mathcal{R}^{d_f}$ and $\mathbf{w}(\vec{r}): \mathcal{R}^3 \rightarrow \mathcal{R}^{d_f} $ be functions with a finite resolution, 
i.e., their Fourier expansion coefficients are zero  for all $l \leq$ than some maximum expansion order $L$.
In order to obtain a vector $\mathbf{h} \in  \mathcal{R}^{d_f}$,  we integrate element-wise product of two functions over the 3D space.
We are using the following relation to compute our vector representation,
 \begin{equation}
 \begin{split}
	\mathbf{h} =  \int_{\mathcal{R}^3} \mathbf{f}(\vec{r})\mathbf{w}(\vec{r}) d \vec{r} 
=
\frac{1}{(2 \pi)^3} \sum_{l = 0}^{L} \sum_{k = -l}^{l} 
 \int_{0}^{\infty} 
 \mathbf{F}^k_l(\rho) \overline{\mathbf{W}}^k_l(\rho)
	 \rho^2 d \rho
,
 \end{split}	
\end{equation}
where we approximate the last integral with a numerical integration. 

\section{Network architectures}

Table \ref{table1} lists three network architectures used in our experiments.

\begin{table}[h]
\centering
\begin{tabular}{ | m{5em} | m{10em} | m{10em} | m{10em} |} 
 \hline
 Layers & Baseline & 6DCNN-1 & 6DCNN-2 \\
\hline
Edges features embedding &Filter $(410, 10)$;  $N_p =$ 4,100&Filter $(410, 10)$; $N_p =$ 4,100 & Filter $(410, 10)$; $N_p =$ 4,100  \\

 \hline
 Nodes features embedding &Filter $(168, 40)$;  $N_p =$ 6,720&Filter $(168, 40)$; $N_p =$ 6,720 & Filter $(168, 40)$; $N_p =$ 6,720  \\ 
 
  \hline
 6D conv layer  & - & - & Filter $ \forall l (l, 4, 40, 40)$; $N_p =$ 131,200\\ 
 \hline
 Message-passing& - & - & No trainable  parameters\\ 
 \hline
 6D conv layer  & - & Filter $ \forall l (l, 4, 40, 40)$; $N_p =$ 131,200 & Filter $ \forall l(l, 4, 40, 40)$; $N_p =$ 131,200\\ 
 \hline
 Continuous to discrete   & Filter $ \forall l (l, 4, 40)$; $N_p =$ 3,200  & Filter $ \forall l(l, 4, 40)$; $N_p =$ 3,200 & Filter $ \forall l (l, 4, 40)$; $N_p =$ 3,200\\ 
 \hline
 GC layer   & Filters  $ (40, 14, 10)$ and $ (40, 14)$; $N_p =$ 6,174  & Filters  $ (40, 14, 10)$ and $ (40, 14)$; $N_p =$ 6,174 & Filters  $ (40, 14, 10)$ and $ (40, 14)$; $N_p =$ 6,174\\ 
  \hline
 GC layer   & Filters  $ (14, 5, 10)$ and $ (14, 5)$; $N_p = 775$  & Filters  $ (14, 5, 10)$ and $ ( 14, 5)$; $N_p = 775$ & Filters  $ (14, 5,  10)$ and $ (14, 5)$; $N_p = 775$\\ 
 \hline
 GC layer   & Filters  $ (5, 1,10)$ and $ (5, 1)$; $N_p = 55$  & Filters  $ (5, 1, 10)$ and $ ( 5,1)$; $N_p = 55$ & Filters  $ (5, 1,  10)$ and $ (5, 1)$; $N_p = 55$\\ 
 \hline
 \hline
 The total number of parameters & 21,026& 187,326& 283,426\\
 \hline
 
\end{tabular}
\caption{Comparison of the 6DCNN and the baseline network architectures with the number of trainable parameters $N_p$ on each layer.}
\label{table1}
\end{table}

\section{Local scores' prediction results.}
\label{app:local_scores}
Table \ref{table:local_scores} lists  Spearman rank correlations of local quality predictions with the corresponding ground-truth values of our three networks and the state-of-the-art methods on model structures of $11$ targets from CASP13. 
 For the comparison, we chose only model structures that had both publicly available target structures and predictions of local scores from other methods.

\begin{table}[ht]
\centering
\begin{tabular}{ | c | c |}
\hline
Methods & $|\rho|$  \\
\hline
VoroCNN&	$\mathbf{0.795}$\\
ModFOLD7&	0.776\\
VoroMQA-A&	0.665\\
ProQ3&	0.744\\
ProQ4&	0.728\\
Ornate&	0.540\\
\hline
Baseline&	0.735\\
6DCNN-1&	0.774\\
6DCNN-2	&0.763\\
\hline
\end{tabular}
\caption{
Spearman rank correlations between local scores and its predictions of the 6DCNN networks, the baseline architecture, and the state-of-the-art methods on the unrefined CASP13 stage2 dataset.
Only 11 targets with the publicly available structure are included here.
The best value is highlighted in bold.
}
\label{table:local_scores}

\end{table}


\end{document}